\documentclass[sigconf,bookmarksnumbered,unicode,bookmarks=false,balance=false,hidelinks,timestamp=false,authorversion,nonacm]{acmart}

\pdfoutput=1
\usepackage[utf8]{inputenc}

\copyrightyear{2024} 
\acmYear{2024} 
\setcopyright{rightsretained} 
\acmConference[ICSE '24]{2024 IEEE/ACM 46th International Conference on Software Engineering}{April 14--20, 2024}{Lisbon, Portugal}
\acmBooktitle{2024 IEEE/ACM 46th International Conference on Software Engineering (ICSE '24), April 14--20, 2024, Lisbon, Portugal}\acmDOI{10.1145/3597503.3639143}
\acmISBN{979-8-4007-0217-4/24/04}


\usepackage[T1]{fontenc}
\usepackage{url}
\usepackage{flushend}
\usepackage{xspace}
\usepackage[skip=5pt]{caption}
\usepackage[inline]{enumitem}
\usepackage{ifthen}
\usepackage{balance}

\usepackage[frozencache,cachedir=minted-cache]{minted}
\setminted{breaklines,breakanywhere}
\setmintedinline{breaklines,breakafter=.}

\usepackage{microtype}
\usepackage{multirow}
\usepackage[]{siunitx}
\newcommand*\np[2][z]{
\ifx z#1%
$\num{#2}$%
\else%
$\qty{#2}{#1}$%
\fi\xspace%
}

\usepackage{fp}
\usepackage{xfp}

\usepackage{graphicx}
\usepackage{subcaption}
\usepackage{pifont}

\usepackage{booktabs}
\usepackage{multirow}
\usepackage{threeparttable}
\usepackage{xcolor}
\usepackage{array,makecell}
\usepackage{diagbox}
\usepackage{tabularx}

\usepackage{mathtools}
\usepackage{amsmath}

\usepackage{tcolorbox}
\usepackage{pgfplots}
\usepackage{pgfplotstable}

\usepackage{enumitem}
\usepackage[framemethod=TikZ]{mdframed}
\usepackage[scaled]{beramono} 

\usepgfplotslibrary{statistics}
\usetikzlibrary{calc,patterns,fpu} 
\pgfplotsset{compat=1.14} 

\pgfplotsset{
    /pgf/declare function={
        Floor(\x) = round(\x-0.49);
    },
    show sum on top/.style={
        /pgfplots/scatter/@post marker code/.append code={%
            \path let \p1=($(normalized axis cs:%
                        \pgfkeysvalueof{/data point/x},%
                        \pgfkeysvalueof{/data point/y})%
                        -(normalized axis cs:\pgfkeysvalueof{/data point/x},0)$)
            in node[
                at={(normalized axis cs:%
                        \pgfkeysvalueof{/data point/x},%
                        \pgfkeysvalueof{/data point/y})%
                },
                anchor={-90*sign(\y1)},yshift={sign(\y1)*2pt}
            ]
            {\pgfmathprintnumber[fixed, precision=1]{\pgfkeysvalueof{/data point/y}}};
        },
    }
}

\definecolor{blau_1a}{RGB}{93,133,195}
\definecolor{blau_2a}{RGB}{0,156,218}
\definecolor{gruen_3a}{RGB}{80,182,149}
\definecolor{gruen_4a}{RGB}{175,204,80}
\definecolor{gruen_5a}{RGB}{221,223,72}
\definecolor{orange_6a}{RGB}{255,224,92}
\definecolor{orange_7a}{RGB}{248,186,60}
\definecolor{rot_8a}{RGB}{238,122,52}
\definecolor{rot_9a}{RGB}{233,80,62}
\definecolor{lila_10a}{RGB}{201,48,142}
\definecolor{lila_11a}{RGB}{128,69,151}

\definecolor{blau_1b}{RGB}{0,90,169}
\definecolor{blau_2b}{RGB}{0,131,204}
\definecolor{gruen_3b}{RGB}{0,157,129}
\definecolor{gruen_4b}{RGB}{153,192,0}
\definecolor{gruen_5b}{RGB}{201,212,0}
\definecolor{orange_6b}{RGB}{253,202,0}
\definecolor{orange_7b}{RGB}{245,163,0}
\definecolor{rot_8b}{RGB}{236,101,0}
\definecolor{rot_9b}{RGB}{230,0,26}
\definecolor{lila_10b}{RGB}{166,0,132}
\definecolor{lila_11b}{RGB}{114,16,133}

\newboolean{showcomments}
\setboolean{showcomments}{true}

\newcommand{\ShowAbsoluteNumber}[1]{%
\ifnum #1<10%
{\hspace*{0pt}#1}%
\else%
\ifnum #1<100%
{\hspace*{0pt}#1}%
\else%
\ifnum #1<1000%
{\hspace*{0pt}#1}%
\else%
{\numprint{#1}}%
\fi%
\fi%
\fi%
}

\newcommand{\ShowPercentage}[2]{%
\FPeval\percentage{round(#1/#2*100,0)}%
\FPeval\percentageOneDecimal{round(#1/#2*100,1)}%
\ifnum \percentage=0%
{\np[\%]{\FPprint{percentageOneDecimal}}}%
\else%
\ifnum \percentage<10%
{\np[\%]{\FPprint{percentageOneDecimal}}}%
\else%
{\np[\%]{\FPprint{percentageOneDecimal}}}%
\fi%
\fi%
\xspace
}
\newlength\BARSIZE  
\newcommand{\inlinechart}[2]{%
\FPeval{\BLACKBARSIZE}{#1/#2}\textcolor{black!80}{\rule{\BLACKBARSIZE\BARSIZE}{1.6ex}}%
\FPeval{\BLACKBARSIZE}{1 - (#1/#2)}\textcolor{black!10}{\rule{\BLACKBARSIZE\BARSIZE}{1.6ex}}%
}

\newcommand*\percent[3][v]{%
\ifx q#1%
    \np{#2}/\np{#3}(\ShowPercentage{#2}{#3})\else%
\ifx s#1%
    \ShowPercentage{#2}{#3}\else%
\ifx p#1%
    \np{#2}(\ShowPercentage{#2}{#3})\else%
\ifx c#1%
    \inlinechart{#2}{#3}%
\else%
    \np{#2}%
    \ifx r#1%
        /\np{#3}%
    \fi%
    \hspace*{0.5ex}(\ShowPercentage{#2}{#3}) %
    \inlinechart{#2}{#3}%
    \xspace
\fi\fi\fi\fi%
}
\input{macros.tex}

\definecolor{mygray}{RGB}{240,240,240}

\newcommand{\answer}[2]{\vspace{.2cm}{\centering\setlength{\fboxrule}{0.1pt}\fbox{\colorbox{mygray}{\parbox{0.95\columnwidth}{\textbf{Answer to RQ#1}. #2}}}\vspace{.2cm}}}

\ifthenelse{\boolean{showcomments}}
 { }
        { }

\newcommand{\code}[2][bash]{\mintinline{#1}{#2}\xspace}

\newcommand{\smell}[1]{\emph{#1}\xspace}

\definecolor{eminence}{RGB}{108,48,130}
\definecolor{weborange}{RGB}{255,165,0}
\definecolor{frenchplum}{RGB}{129,20,82}
\definecolor{darkgreen}{RGB}{10, 92, 10}

\mdfdefinestyle{mpdframe}{
    nobreak                     =true,
    backgroundcolor             =black!3,
    frametitlebackgroundcolor   =black!15,
    frametitlerule              =false,
    roundcorner                 =5pt,
    innermargin                 =0.3cm,
    innerleftmargin             =0.3cm,
    innerrightmargin            =0.3cm,
    innertopmargin              =0.3cm,
    innerbottommargin           =0.3cm,
    shadowsize                  =1pt
}

\title{Empirical Study of the Docker Smells Impact on the Image Size}
\author{Thomas Durieux}
\affiliation{%
    \institution{TU Delft}
    \country{The Netherlands}
}
\email{thomas@durieux.me}
\orcid{0000-0002-1996-6134}

\begin{document}
\begin{abstract}
Docker, a widely adopted tool for packaging and deploying applications leverages Dockerfiles to build images. However, creating an optimal Dockerfile can be challenging, often leading to ``Docker smells'' or deviations from best practices. This paper presents a study of the impact of \np{\nbRule} Docker smells on the size of Docker images. 

To assess the size impact of Docker smells, we identified and repaired \np{\nbRepairedSmellInBuildable} Docker smells from \np{\countBuildedDockerfile} open-source Dockerfiles.
We observe that the smells result in an average increase of \avgSpaceSaved (\np[\%]{4.6}) per smelly image.
Depending on the smell type, the size increase can be up to \np[\%]{10}, and for some specific cases, the smells can represent \np[\%]{89} of the image size.
Interestingly, the most impactful smells are related to package managers which are commonly encountered and are relatively easy to fix.

To collect the perspective of the developers regarding the size impact of the Docker smells, we submitted \np{\nbPRs} pull requests that repair the smells and we reported their impact on the Docker image to the developers. 
\percent[q]{\nbAcceptedPR}{\nbPRs} of the pull requests have been merged and they contribute to a saving of \totalSpaceMerged (\percent[s]{3.46}{21.08}). 
The developer's comments demonstrate a positive interest in addressing those Docker smells even when the pull requests have been rejected.
\end{abstract}

\maketitle

\section{Introduction}

Docker is a widely adopted tool among developers and organizations for packaging, deploying, and running applications in lightweight, portable containers. A critical component of Docker is the Dockerfile, a straightforward text file based on shell that outlines the necessary steps to build a Docker image. However, creating an optimal Dockerfile can be challenging, particularly when shell best practices differ from the ones in Docker.
When there is a deviation from these best practices, we refer to it as a ``Docker smell''. Docker smells are commonly found within Dockerfiles because many developers who create them may lack expertise in this area \cite{10.1145/3377811.3380406}. Furthermore, the best practices used in interactive shells often contrast with those applicable to shells within Dockerfiles, resulting in suboptimal Docker images.

Previous research conducted by academics and industry has primarily focused on detecting Docker smells. 
Several linters, such as \textit{Binnacle} \cite{10.1145/3377811.3380406}, \textit{hadolint} \cite{hadolint}, \textit{dockerfilelint} \cite{dockerfilelint}, \textit{docker-bench-security} \cite{docker-bench-security}, and \textit{dockle} \cite{dockle}, have been developed specifically to identify a wide range of Docker smells. 
However, these tools suffer from limited recognition in the developer community as multiple studies show the almost systematic presence of smells in Dockerfiles \cite{9240654,cito2017empirical}. 
One possible reason for the lack of recognition among developers may be the absence of studies on the impact of these smells, making it challenging to justify investing effort into addressing them.

In this contribution, we aim to address this specific problem by investigating the impact of Docker smells on the image size.
We focus on the image size since it impacts multiple aspects of the Docker ecosystem.
Firstly, the image size impacts the Docker images selection by the developers, smaller images have more chances to be selected~\cite{10.1145/3603111}.
It also contributes to the size of the Docker registry which was already reaching \np[PB]{1} in 2019 \cite{zhao2019large} for public repositories and is expected to be much bigger for private repositories.
It also impacts the download latency of the Docker images \cite{zhao2020duphunter} which is problematic in large deployment environments, as well as increases the attack surface of the Docker images.

In this study, we investigate the size impact of \np{\nbRule} Docker smell types originally identified by Henkel et al. \cite{10.1145/3377811.3380406} as having a potential impact on image size.
The size impact is measured by identifying and removing \np{\nbRepairedSmellInBuildable} real Docker smells from \np{\countBuildedDockerfile} open-source Dockerfiles.
We then investigate the developers' perspectives and interests in those Docker smells in order to identify if notifying the developers about those smells is relevant or not. 
This aspect has been performed by opening \np{\nbPRs} pull requests that repair and report the impact of \np{\nbMergedRepairedSmell} Docker smells.
The detection and repair are performed by our tool, \tool, which has been specifically developed for this purpose.

Our observations reveal that Docker smells exert a substantial impact on the size of Docker images. On average, the Docker smells lead to a size increase of \avgSpaceSaved or \np[\%]{4.6} per image. Additionally, this bloat translates to a total additional \totalSpaceSavedPerWeek in transferred data per week on DockerHub. Notably, we found that the most impactful smells identified in this study are associated with the utilization of package manager commands.
Those smells also happen to be among the most frequently encountered ones, which means that identifying and repairing a few smells can have a huge impact and improve the quality of the Docker images.

As direct evidence of the relevance of smells repair, \percent[q]{\nbAcceptedPR}{\nbPRs} of pull requests have been successfully merged, indicating developers' interest in repairing Docker smells, \np{\nbPendingPR} pull requests are still waiting for an answer and \np{\nbRefusedPR} pull requests have been rejected because the proposed changes were already included in the repository.
The merged pull requests contribute to a saving of \totalSpaceMerged (\percent[s]{3.46}{21.08}).

In summary, the contributions of this paper include:
\begin{itemize}
\item An empirical study on the impact of Docker smells on the image size,
\item A new dataset of \np{\nbParfumDockerfiles} Dockerfiles extracted from GitHub and a ground truth dataset of \np{\nbDockerfilesGroundTruth} Dockerfiles,
\item \np{\nbPRs} pull requests that repair \np{\nbMergedRepairedSmell} Dockerfile smells,
\item \tool, a tool that detects and repairs automatically \np{\nbRule} types of Docker smells.
\end{itemize}

We are pleased to announce that we have made the results of our study accessible at \cite{repo-expe}. Additionally, the smell detection and repair technique is available at \cite{repo} and can be tested at \url{\urlDemo}.

\section{Background}\label{sec:background}

In this section, we provide the key concepts and background information required for our study.

\textbf{Containers} are a form of virtualization technology designed to offer a more efficient and streamlined approach to software deployment. Unlike traditional virtual machines, containers encapsulate applications and their dependencies, ensuring consistency across different environments. 
By doing so, they enhance portability and facilitate the seamless movement of applications between development, testing, and production environments. 
Containers gained popularity also due to their lower overhead compared to virtual machines \cite{adams2006comparison,kumar2016economically}.

\textbf{Docker} is the most popular container platform that can create, deploy, and run containerized applications.\footnote{Docker: \url{https://www.docker.com}} Docker also has its own Docker registry which is the most popular registry for open-source Docker images.

\textbf{Docker Image} is an executable package for Docker that includes everything needed to run a piece of software, including the code, a runtime, libraries, environment variables, and config files. Docker images are built using instructions contained in a Dockerfile.

\textbf{Dockerfile} is a text file that contains instructions for building a Docker image. 
The instructions define the base image to use (\code[dockerfile]{FROM <image>}), the files to include (\code[dockerfile]{COPY <source> <dest>}), the ports to open (\code[dockerfile]{PORT <port>}), the entry point (\code[dockerfile]{ENTRYPOINT <script>}), and the scripts to execute (\code[dockerfile]{RUN <script>}).
The scripts declared in the Dockerfiles define the actions that need to be performed to create the Docker image.
Those scripts are shell commands which are generally bash or PowerShell (for Windows Docker image).

\textbf{Docker Smell} refers to a potential issue, problem, or suboptimal configuration with a Dockerfile or Docker image \cite{wu2020characterizing}. 
This issue is generally detected when the Dockerfile or image violates some best practices.
Common Docker smells include bloated images, misconfiguration, misuse of commands, and security issues. 
Identifying and addressing these smells can help improve the efficiency, security, and maintainability of a Docker-based project \cite{wu2020characterizing}.
In this paper, we focus on smells inside the Dockerfiles.

\textbf{Binnacle} by Henkel et al. \cite{10.1145/3377811.3380406} is a tool that studies and detects Docker smells in Dockerfiles. 
The particularity of this work compared to other linters such as Hadolint is that it not only detects Docker smells but also analyzes the presence of those smells inside GitHub and compares it to a high-quality set of Dockerfiles.
Additionally, it also categorizes the impacts of the smells they observed. 
Interestingly, the majority of the smells are related to space waste; this observation initiated this study to measure the actual impact of those smells on the image size.
The Docker smells reported by Binnacle have a small overlap with other existing linters such as Hadolint which only supports \np{4} smell types, also supported by Binnacle, that impact image size.
Indeed, most of the Binnacle smells are related to the shell while Hadolint focused on the Docker instructions and the size impact is related to the shell usage.

\section{Methodology}

We describe the empirical study we conduct on the impact of Docker smells.
We first present the methodology that we follow to perform this empirical study.
Then, we present the datasets that we use for the empirical study.
We follow by describing how we detect and repair Docker smells with our tool called \tool.

\subsection{Methodology Overview}
In order to measure the impact of the smells on image size.
We follow the following methodology for each Dockerfile.
First, we identify the smells that are present inside the Dockerfiles.
If a smell is present, we build the smelly Dockerfile to produce a Docker image and we measure the size of that image.
We then repair the smell and produce a new Dockerfile without smells.
We then build the repaired Dockerfile and measure the size of the new image.
Finally, the difference in size between the original and the repaired image is the impact of the detected smells.

\subsection{Research Questions}
In this section, we present the impact of Docker smells on Docker image size. 
We design and conduct an empirical evaluation to answer the following research questions:

\begin{itemize}
\item[RQ1] \textbf{What is the effectiveness of our approach in detecting and repairing Docker smells?}
This first research question aims to validate a crucial aspect of our methodology: being able to detect and repair Docker smells.
To do so, we first measure the effectiveness of smell detection on a ground truth dataset.
Then, we conduct a quantitative analysis of the repair of \np{\nbTotalDockerfilesWithSmell} Dockerfiles that contain at least one smell. 
Finally, we selected \np{\countBuildedDockerfile} smelly Dockerfiles and built them to ensure that the repairs do not break the Docker builds.
\item[RQ2] \textbf{What is the impact of the identified Docker smells on the Docker image size?}
In this second research question, we study the impact of the Docker smell on the size of the Docker images.
To answer this question, we measure the image size before and after the repair for the \np{\countBuildableDockerfileAfterRepair} Dockerfiles. 
We also study which smells have the most impact the most the Docker image size and the effect of the smells on the DockerHub bandwidth.
Finally, we measure the impact of the smells in terms of bandwidth on Dockerhub. 
\item[RQ3] \textbf{What is the developers' attitude towards Docker smells?}
In the final research question, we aim to evaluate the interest that the developers have in the repair of Docker smells impacting image size.
To do so, we opened \np{\nbPRs} pull requests that fix the identified smells and we analyzed the responses of the developers.
\end{itemize}

By addressing these research questions, we aim to analyze the Docker smells impact on image size and developers' attitudes regarding these smells.

\subsection{Docker Smells}\label{sec:smell_rules}
As previously mentioned, for this study we focus on the Docker smell that introduces an increase in size as presented by Henkel et al.~\cite{10.1145/3377811.3380406}.
We therefore ignore the smells that are related to the security or build reliability.
During this study, we will therefore focus on \np{\nbRule} smells.
\autoref{tab:smells} describes the smells and provides the ID that we will use to refer to them.
Additionally, the table includes the results of our first research question which we will present later on. 

\begin{table*}[ht]
    \centering
    \caption{The considered Docker smells and the detection rate by \tool and Binncale in our Ground Truth dataset.}\label{tab:smells}
    \begin{tabularx}{0.95\linewidth}{@{}r l X| rr@{}}
\toprule
\# & Smell ID & Smell Description & \tool & Binnacle\\
\midrule
1  & pipUseCacheDir            & Clean cache after \code{pip install}. & \percent[q]{82}{82} & \percent[q]{67}{82}\\
2  & npmCacheCleanUseForce     & Clean cache after \code{npm install}. & \percent[q]{2}{2} & \percent[q]{2}{2}\\
3  & mkdirUsrSrcThenRemove     & Remove /usr/src/* after usage. & - & - \\
4  & rmRecurisveAfterMktempD   & Remove temporary folders.  & - & - \\
5  & tarSomethingRmTheSomething& Remove tar files after decompression. & \percent[q]{13}{12} & \percent[q]{7}{12}\\
6  & apkAddUseNoCache          & Use \code{--no-cache} flag with \code{apk add}. & \percent[q]{8}{8} & \percent[q]{8}{8}\\
7  & aptGetInstallUseNoRec     & Use \code{--no-install-recommends} flag in \code{apt-get install}. & \percent[q]{159}{159} & \percent[q]{122}{159}\\
8  & aptGetInstallRmAptLists   & Remove \code{/var/lib/apt/lists/*} after \code{apt-get install}. &  \percent[q]{153}{153} & \percent[q]{117}{117}\\
9  & gpgVerifyAscRmAsc         & Remove .asc file after usage.  & - & - \\
10 & npmCacheCleanAfterInstall & Force to clean cache after \code{npm install}. & \percent[q]{30}{30} & \percent[q]{28}{28}\\
11 & gemUpdateSystemRmRootGem  & Clean cache after \code{gem update --system}. & \percent[q]{1}{1} & \percent[q]{1}{1}\\
12 & gemUpdateNoDocument       & Add \code{--no-document} flag to the .gemrc config file. & \percent[q]{1}{1} & \percent[q]{1}{1}\\
13 & yumInstallRmVarCacheYum   & Clean cache after \code{yum install}. & \percent[q]{17}{17} & \percent[q]{17}{17}\\
14 & yarnCacheCleanAfterInstall& Clean cache after \code{yarn install}. & \percent[q]{3}{3} & \percent[q]{0}{3}\\
\bottomrule
\end{tabularx}
\end{table*}

\subsection{Datasets}\label{sec:dataset}

In order to study the impact of the smells we had to select and create a new dataset of Dockerfiles. This section will present the dataset that we use in this study.
\autoref{tab:dataset} gives an overview of the main characteristics of our datasets and highlights some of their main differences.

\subsubsection{Ground Truth Dataset}\label{sec:ground_truth}

The second dataset that we consider in this study is a dataset of \np{\nbDockerfilesGroundTruth} unique Dockerfiles. 
That is used to measure the effectiveness of our approach to detect Docker Smells.
The Dockerfiles have been manually annotated to identify the Docker smells.
We randomly selected those Dockerfiles from the Binnacle Dataset. 
We chose a sample size of \np{\nbDockerfilesGroundTruth} Dockerfiles to obtain a dataset that is representative of the Binnacle dataset with a confidence level of \np[\%]{95} and with a margin of error of \np[\%]{5} according to the Cochran’s Sample Size Formula: $n = \frac{z^2 \cdot p(1-p)}{\epsilon^2}$, where $n$ is the required sample size, $z$ is the Z-score corresponding to the desired confidence level (e.g., 1.96 for a \np[\%]{95} confidence level), $p$ is the estimated proportion of the population with a certain characteristic, and $\epsilon$ is the desired margin of error \cite{cochran1977sampling}.
The ground truth dataset is also available on our online artifact \cite{repo-expe}.
We observe that the distribution of the number of instructions in the ground truth dataset and the Binnacle dataset are similar and therefore we are confident that this annotated dataset is representative. 

The methodology to create this dataset is as follows: 1. The authors read the description of the smells to have a clear understanding of the smells.
2. We create a dashboard that displays and annotates the Dockerfiles; the goal is to minimize the effort of the annotation and to focus on the manual detection of the smells.
3. Multiple interactions have been performed to ensure that all smells are identified.
4. As a final check, we carefully analyze the results of Binnacle and ours to identify cases that could have been mislabeled.

At the end of this process, \np{152} Dockerfiles have at least one smell, and in total \np{468} smells have been annotated.
This dataset is as far as we know the first ground truth dataset for Docker smells.

\subsubsection{Binnacle Dataset}
The first dataset that we use is the dataset of unique Dockerfiles presented in the Binnacle paper \cite{10.1145/3377811.3380406}, which contains \np{\nbDockerfilesBinnacle} Dockerfiles extracted from GitHub repositories in 2020. 
The main purpose of this dataset is to compare our smell detection to the baseline: Binnacle.

\subsubsection{\tool Dataset}
The third and final dataset contains \np{\nbParfumDockerfiles} Dockerfiles that were extracted from GitHub repositories in 2022. 
This dataset is used to study the impact of the smells in RQ2 and identify projects where we submit pull requests in RQ3.
We could not use the Binnacle dataset for RQ2 and RQ3 because we could not identify from which repository the Dockerfiles from the Binnacle dataset were and therefore we could not build the Docker images to measure their size nor open pull requests.
To avoid this problem in the future, we include in our dataset the origin repository, commit SHA, and the path of the Dockerfile.

The methodology for creating this new dataset is described in the following.
The first step is to identify an initial set of GitHub repositories. 
We decided to select repositories that are 1) not forks, 2) have at least 10 stars, and 3) have at least 50 commits.
We choose those criteria to obtain Dockerfiles from repositories that have a minimum of activity and that are more likely to have been maintained. 
We ended up with a list of \np{500108} potential repositories.\footnote{Downloaded on July 12, 2022 from \url{https://seart-ghs.si.usi.ch/}}

The next step is to download the file list from the default branch of the latest commit for each repository. 
We were able to download the file list for \np{500022} repositories, the missing file lists are due to unreachable repositories.

The following step is to identify and download the Dockerfiles stored in these repositories. We iterated over the list of files and considered any files that contained the string ``Dockerfile'' (case sensitive) as potential Dockerfiles. 
Finally, we identified the unique Dockerfiles that we use in this study.
This resulted in a collection of \np{\nbParfumDockerfiles} Dockerfiles that constitute the new dataset which is available on our online artifact \cite{repo-expe} as well as the scripts that are used to generate the dataset. 

\begin{table}[t]
    \caption{Characteristics of Binnacle \cite{10.1145/3377811.3380406}, \tool, and Ground Truth  datasets.}
    \label{tab:dataset}
    \centering
    \begin{tabular}{@{}l|r r r@{}}
\toprule
Metric               & Binnacle & \tool & Ground Truth\\
\midrule
Creation date        & 2020 & July 2022  & July 2022 \\
\# Dockerfile        & \np{\nbDockerfilesBinnacle} & \np{\nbParfumDockerfiles} & \np{\nbDockerfilesGroundTruth} \\
\# Smelly Dockerfile & \np{\totalSmellyDockerFileByBinnacleInBinnacle} & \np{\nbParfumDockerfilesWithSmell}  & \np{\nbDockerfilesGroundTruthWithSmell} \\
Total \# Instruction & \np{2223139}                & \np{3637952} & \np{4938}\\
Avg. \# Instruction  & \np{12.45}                  & \np{18.04} & \np{12.86}\\
Med. \# Instruction  & \np{9}                      & \np{12} & \np{9}\\
\bottomrule
    \end{tabular}
\end{table}

\subsection{\tool}\label{sec:tool}
In this section, we present, \tool, a tool we use to detect and repair Docker smells. 
\tool is available on GitHub \cite{repo} and it also has been ported to a browser version which is available at \url{\urlDemo}.

\tool detection of smells is inspected by Binnacle \cite{10.1145/3377811.3380406} and supports the smells that Binnacle reports as being related to space waste.
The major difference between Binnacle and \tool is that \tool repairs those smells and it also links the smells to an AST node which allows much more precise analysis and extensions.

\subsubsection{\tool Steps}

In this section, we briefly explain how \tool works by presenting the six main steps.

\begin{enumerate}
  \item \textbf{Parsing Dockerfile AST}: The first step of {\tool} is to parse the Abstract Syntax Tree (AST) representation of the Dockerfile.
  
  \item \textbf{Parsing shell commands}: \tool parses each Docker command that includes a shell command, i.e., \code[dockerfile]{RUN <cmd>} and compiles it with the Dockerfile AST to form a unified AST.
  
  \item \textbf{Enriching the Docker AST}: Next, \tool enriches the Docker AST by incorporating structural information from the command lines. For instance, consider the command \code[dockerfile]{RUN apt-get install wget} and its AST representation. The enriched AST contains annotations specifying that \code{apt-get} is used to \code{install} packages, and the installed package is \code{wget}. These annotations are added to the corresponding nodes in the AST, highlighting their roles and relationships and they can be used later on by the smell analyzer.
  {\tool} supports a total of \np{\nbSupportedCmd} command lines, which account for \np[\%]{89.05} of all the commands found in the Dockerfiles within our dataset. 
  The remaining commands consist of either custom or infrequent commands. Consequently, these commands will not be part of Docker smells by nature.
  
  \item \textbf{Enriching embedded commands}: \tool enriches commands that are embedded within other commands. For example, the command \code{sudo apt update} contains a main command (\code{sudo}) and an embedded command (\code{apt update}).

  \item \textbf{Detecting Docker smells}: The detection of smells is made by querying the AST. 
  Each smell is associated with an AST query. The detection of smells is detailed in \autoref{sec:smell_detection}.
  
  \item \textbf{Repairing Docker smells}: Once Docker smells are detected, \tool can proceed with the repair. 
  We employ a template-based approach to repair the smells, the details of the repair are available in \autoref{sec:smell_repair}.
\end{enumerate}

\begin{figure*}
\centering
\begin{subfigure}[b]{0.29\textwidth}\vspace{0pt}%
\begin{minted}{typescript}
{
 // look for `npm cache clean`
 query: Q("NPM-CACHE-CLEAN"), 
 consequent: {
  // look for `--force` flag
  inNode: Q("NPM-F-FORCE") } 
}
\end{minted}
\caption{Detect smell.}
\label{lst:query}
\end{subfigure}\hfill%
\begin{subfigure}[b]{0.42\textwidth}%
\begin{minted}{typescript}
function repair(node) {
  // insert --force flag
  node.addChild(BashCommandArgs().addChild(
    BashLiteral("--force")
  ));
}
\end{minted}
\caption{Repair smell.}
\label{lst:repair}
\end{subfigure}\hfill
\begin{subfigure}[b]{0.27\textwidth}%
\begin{minted}{diff}
@@ -21,1 +21,1 @@
-RUN npm cache clean
+RUN npm cache clean --force
\end{minted}
\caption{Generated Dockerfile patch.}
\label{lst:diff}
\end{subfigure}
\end{figure*}

\subsubsection{Smell Detection}\label{sec:smell_detection}

The smell detection of \tool uses a template matching system to identify patterns inside the Dockerfile AST. 
In total, \tool supports \np{32} Docker smell detections, but we only considered the \np{\nbRule} that are related to space waste. 
The list of the \np{32} supported smells is presented in our repository \cite{repo}, even if they are not the focus of this paper, developers can still use \tool to detect and fix them.

The considered smells are described in \autoref{sec:smell_rules}.
Each rule is defined as a query, specifying the required AST nodes that need to be present to trigger the smell.
An additional post-condition specifies additional AST nodes that should be present before, after, or inside the matched node.
\autoref{lst:query} presents an example of such a template matching. In this example, we detect that the flag \code{-f} is missing within the command \code{npm cache clean}.
In this example, we look for the command \code{npm cache clean} using the query \code[typescript]{Q("NPM-CACHE-CLEAN")}. 
The post-condition verifies that the flag \code{-f} is not present inside the node with the query \code[typescript]{Q("NPM-F-FORCE")}.  
If those two queries have a match, \tool has detected the smell and it is reported to the developer.

\subsubsection{Smell Repair}\label{sec:smell_repair}
Once a smell is detected, \tool repairs the Dockerfile by modifying its AST.
This is a novelty of \tool, as far as we know \tool is the first tool that fixes smells in Dockerfiles.
\autoref{lst:repair} presents an example of how the \tool modifies the AST to fix the smell. 
In this particular example, the smell is related to the command \code{npm cache clean}. \tool repairs the smell by adding the \code{--force} flag as an argument to the \code{npm cache clean} command.
Once the AST is transformed, the detection of the smell is triggered again to verify that the repair was made properly. If the smell is still detected, the repair is rollback to avoid introducing inappropriate changes to the Dockerfiles.

After modifying the AST, \tool can reprint the AST into a Dockerfile.
The reprinting process in \tool utilizes a pretty-print feature, resulting in the reprinted AST containing only the modified nodes. This approach minimizes the changes made to the Dockerfile while addressing the detected smells. An example of such transformation can be seen in \autoref{lst:diff}, showcasing the differences between the original Dockerfile and the repaired version.

\section{Study Results}
In this section, we present and discuss the answers to our research questions.

\subsection{RQ1: Smell Detection \& Repair Effectiveness}\label{sec:rq1}

In this first research question, we assess the effectiveness of our methodology in identifying and repairing Docker smells. 
The detection and repair are handled by our tool: \tool and consequently, we will also evaluate the effectiveness of our tool.
To simplify the narrative, we will refer to our approach as \tool in this research question.

This evaluation is divided into two parts: first, we evaluate the effectiveness of detecting the smells by analyzing the smell detection rate of our approach on the ground truth dataset and also comparing it to the baseline: Binnacle~\cite{10.1145/3377811.3380406}.
Second, we evaluate the repair effectiveness of our approach and its impact on the Docker build, i.e., build failure rate.

\subsubsection{\tool vs Binnacle} \label{sec:ground_truth_eval}

To increase our confidence in our approach, we measure our approach detection rate and compare it to the baseline, Binnacle, on our ground truth dataset (see \autoref{sec:ground_truth}).
\autoref{tab:smells} presents the detection rate of \tool and Binnacle.
We observe that \tool has almost a perfect detection rate.
Only in one case, \tool produces a false positive for the smell \smell{tarSomethingRmTheSomething} while Binnacle produces at least 89 false negatives.
We cannot get the precise rate because Binnacle only reports the number of each detected smell without their position which could lead to an unprecise comparison with the ground truth.

\autoref{lst:parfum_false_positive} presents the only false positive reported by \tool. It happens because \tool does not succeed in identifying that the developers already removed the \texttt{tar} using the command: \code{rm -rf /tmp/firefox.*}.

Based on those results, we can be confident in the detection effectiveness of our approach.

\begin{listing}[t]
\begin{minted}{dockerfile}
RUN FIREFOX_URL="https://download.mozilla.org/?product=firefox-latest-ssl&os=linux64&lang=en-US"
  && ACTUAL_URL=$(curl -Ls -o /dev/null -w %{url_effective} $FIREFOX_URL)
  && curl --silent --show-error --location --fail --retry 3 --output /tmp/firefox.tar.bz2 $ACTUAL_URL
  && sudo tar -xvjf /tmp/firefox.tar.bz2 -C /opt
  && sudo ln -s /opt/firefox/firefox /usr/local/bin/firefox
  && sudo apt-get install -y libgtk3.0-cil-dev libasound2 libasound2 libdbus-glib-1-2 libdbus-1-3
  && rm -rf /tmp/firefox.*
  && firefox --version
\end{minted}
\caption{False positive produced by \tool. \tool did not identify that \texttt{firefox.tar.bz2} was removed by \code{rm -rf /tmp/firefox.*} commands and therefore identifies the \smell{tarSomethingRmTheSomething} smell in this snippet. }
\label{lst:parfum_false_positive}
\end{listing}

\subsubsection{\tool Repair Effectiveness}\label{sec:repair_effectiveness}

\begin{table*}[t]
    \caption{The occurrence of each smell before and after the repair using \tool on Binnacle and \tool datasets.}
    \label{tab:nbViolation}
    \centering
    \begin{tabular}{@{}l r r|r r@{}}
    \toprule
\multirow{2}{*}{Docker Smell} &  \multicolumn{2}{c|}{\# Docker Smell}  & \multicolumn{2}{c}{\# Dockerfile with Smell} \\
&  Before Repair &  After Repaired  &  Before Repair &  After Repaired\\\midrule
pipUseNoCacheDir & \percent{76856}{\nbSmell} & \percent{7}{\nbSmellAfterRepair} & \percent{41282}{\nbTotalDockerfilesWithSmell} & \percent{3}{\nbTotalRepairedDockerfilesWithSmell} \\
npmCacheCleanUseForce & \percent{2447}{\nbSmell} & \percent{6}{\nbSmellAfterRepair} & \percent{2413}{\nbTotalDockerfilesWithSmell} & \percent{6}{\nbTotalRepairedDockerfilesWithSmell} \\
mkdirUsrSrcThenRemove & \percent{4777}{\nbSmell} & \percent{28}{\nbSmellAfterRepair} & \percent{4329}{\nbTotalDockerfilesWithSmell} & \percent{27}{\nbTotalRepairedDockerfilesWithSmell} \\
rmRecursiveAfterMktempD & \percent{768}{\nbSmell} & \percent{11}{\nbSmellAfterRepair} & \percent{491}{\nbTotalDockerfilesWithSmell} & \percent{11}{\nbTotalRepairedDockerfilesWithSmell} \\
tarSomethingRmTheSomething & \percent{20902}{\nbSmell} & \percent{129}{\nbSmellAfterRepair} & \percent{14660}{\nbTotalDockerfilesWithSmell} & \percent{97}{\nbTotalRepairedDockerfilesWithSmell} \\
apkAddUseNoCache & \percent{15094}{\nbSmell} & \percent{0}{\nbSmellAfterRepair} & \percent{11671}{\nbTotalDockerfilesWithSmell} & \percent{0}{\nbTotalRepairedDockerfilesWithSmell} \\
aptGetInstallUseNoRec & \percent{172028}{\nbSmell} & \percent{1}{\nbSmellAfterRepair} & \percent{81448}{\nbTotalDockerfilesWithSmell} & \percent{1}{\nbTotalRepairedDockerfilesWithSmell} \\
aptGetInstallThenRemoveAptLists & \percent{142187}{\nbSmell} & \percent{389}{\nbSmellAfterRepair} & \percent{74958}{\nbTotalDockerfilesWithSmell} & \percent{209}{\nbTotalRepairedDockerfilesWithSmell} \\
gpgVerifyAscRmAsc & \percent{157}{\nbSmell} & \percent{0}{\nbSmellAfterRepair} & \percent{144}{\nbTotalDockerfilesWithSmell} & \percent{0}{\nbTotalRepairedDockerfilesWithSmell} \\
npmCacheCleanAfterInstall & \percent{32437}{\nbSmell} & \percent{63}{\nbSmellAfterRepair} & \percent{24248}{\nbTotalDockerfilesWithSmell} & \percent{52}{\nbTotalRepairedDockerfilesWithSmell} \\
gemUpdateSystemRmRootGem & \percent{505}{\nbSmell} & \percent{2}{\nbSmellAfterRepair} & \percent{457}{\nbTotalDockerfilesWithSmell} & \percent{2}{\nbTotalRepairedDockerfilesWithSmell} \\
gemUpdateNoDocument & \percent{390}{\nbSmell} & \percent{0}{\nbSmellAfterRepair} & \percent{345}{\nbTotalDockerfilesWithSmell} & \percent{0}{\nbTotalRepairedDockerfilesWithSmell} \\
yumInstallRmVarCacheYum & \percent{24124}{\nbSmell} & \percent{95}{\nbSmellAfterRepair} & \percent{12083}{\nbTotalDockerfilesWithSmell} & \percent{53}{\nbTotalRepairedDockerfilesWithSmell} \\
yarnCacheCleanAfterInstall & \percent{5010}{\nbSmell} & \percent{12}{\nbSmellAfterRepair} & \percent{4041}{\nbTotalDockerfilesWithSmell} & \percent{12}{\nbTotalRepairedDockerfilesWithSmell} \\
\midrule
Total & \np{\nbSmell} & \np{\nbSmellAfterRepair} & \np{\nbTotalDockerfilesWithSmell} & \np{\nbTotalRepairedDockerfilesWithSmell} \\
\bottomrule
    \end{tabular}

\end{table*}
\begin{table}[t]
    \caption{The number of build errors per smell.}
    \label{tab:build_errors_per_rule}
    \centering
    \begin{tabular}{@{}l r@{}}
    \toprule
Docker Smell &  \# Build Errors \\
\midrule
aptGetInstallUseNoRec & \percent{312}{\countBuildError} \\
aptGetInstallThenRemoveAptLists & \percent{254}{\countBuildError} \\
pipUseNoCacheDir & \percent{115}{\countBuildError} \\
npmCacheCleanAfterInstall & \percent{32}{\countBuildError} \\
tarSomethingRmTheSomething & \percent{48}{\countBuildError} \\
apkAddUseNoCache & \percent{17}{\countBuildError} \\
mkdirUsrSrcThenRemove & \percent{6}{\countBuildError} \\
yumInstallRmVarCacheYum & \percent{4}{\countBuildError} \\
gemUpdateSystemRmRootGem & \percent{1}{\countBuildError} \\
gemUpdateNoDocument & \percent{1}{\countBuildError} \\
npmCacheCleanUseForce & \percent{1}{\countBuildError} \\
\bottomrule
    \end{tabular}
\end{table}

We are now looking at the effectiveness of \tool to repair the Docker smells.
As far as we know no tool or dataset could be used to compare the results of \tool. 
Therefore, we use \tool to automatically repair the smells in the $ \nbBinnacleDockerfiles (Binnacle~Dataset) + \nbParfumDockerfiles (Parfum~Dataset) = \nbTotalDockerfiles$ Dockerfiles, \percent[p]{\nbTotalDockerfilesWithSmell}{\nbTotalDockerfiles} of them contain at least one smell.
We then verified that the smells were fixed by analyzing the repaired Dockerfiles.
Due to the high detection rate presented in the first part of this research question, we can be confident about the repair rate.
To increase our confidence in checking if \tool is not breaking builds, we built the Docker image for a selection of Dockerfiles to ensure that the repair did not break the Docker build.
This will give us some indications of the reliability of the repair. We do know that it is not a perfect oracle and it does not guarantee that the behavior of the images is preserved. However, we could not identify a way that would allow us to verify the behavior of Docker images at a meaningful scale.

\autoref{tab:nbViolation} presents the results of the smell repairs. 
The first column of \autoref{tab:nbViolation} contains the name of the smell, and the second column contains the number of occurrences of this smell.
The third column contains this information after the repair.
The fourth and fifth columns contain the same information but instead count the number of Dockerfiles, i.e., a Dockerfile can contain more than one occurrence of a specific smell.
The results show that \tool is able to repair \percent[p]{\nbRepairedSmell}{\nbSmell} Docker smells. 
The smell \smell{aptGetInstallThenRemoveAptLists} is the smell that is the most present after repair followed by \smell{tarSomethingRmTheSomething} and \smell{yumInstallRmVarCacheYum}.
However, those cases are rare and should not impact significantly the results of our study.

We now verify that \tool does not break the build.
To do so, we build the Docker images before and after the repair.
We could not scale the build of the \np{\nbTotalDockerfilesWithSmell} Dockerfiles due to the amount of computing it would have required.
Indeed, it takes on average \avgExecutionTime to build a Docker image. 
Additionally, the rate-limited imposed by Dockerhub would also block to perform this experiment on all the images.
Instead, we selected all the Dockerfiles that are located at the root of the repositories, that are exactly named \code{Dockerfile}, and that contain at least one smell.
We chose those criteria because we expect that those Dockerfiles are the main Dockerfiles of the repositories.
We end up with \np{\countBuildedDockerfile} Dockerfiles and we only succeeded in building \percent[p]{\countBuildableDockerfile}{\countBuildedDockerfile} of them which illustrates the complexity of reproducing Docker builds.

Once, we identify the \np{\countBuildableDockerfile} Dockerfiles that are buildable and apply \tool on them, and proceed to rebuild the Dockerfiles after the repair. \np{\countBuildableDockerfileAfterRepair} Dockerfiles build after the repair which results in \percent[p]{\countBuildError}{\countBuildableDockerfile} build failures (or build flakiness). 
It is difficult to estimate the number of builds that are failing due to the build flakiness.
However, it is reasonable to believe that the majority is due to \tool repairs.

\autoref{tab:build_errors_per_rule} presents the number of builds that finish with an error per smell type. 
Note that we consider that all applied repairs have impacted the build status.
We observe that the vast majority of the errors are related to the rules \smell{aptGetInstallUseNoRec} and \smell{aptGetInstallThenRemoveAptLists}.
Those rules can break builds when a recommended package is removed when it is required or when \tool removes the cache when it was already empty.

In a few cases, \tool produces invalid repairs such as \autoref{lst:invalid_repair}. 
In this case, \tool places \code{rm gsl.tgz} after the change of directory (\code{cd gsl-1.16}), the file \code{gsl.tgz} is, therefore, not found and the build fails.

\begin{listing}[t]
\begin{minted}{dockerfile}
RUN wget -O gsl.tgz ftp://ftp.gnu.org/gsl-1.16.tar
  && tar -zxf gsl.tgz && mkdir gsl
  && cd gsl-1.16 && ./configure --prefix=/app/gsl
  && make && make install
  && rm gsl.tgz                      # Added line
\end{minted}
\caption{Example of invalid repair made by \tool for the repository \href{https://github.com/olavolav/te-causality}{github.com/olavolav/te-causality}.}
\label{lst:invalid_repair}
\end{listing}

\answer{1}{
We show that our approach is able to detect all the Docker smells in our ground truth dataset with only one false positive while also being able to repair \percent[s]{\nbRepairedSmell}{\nbSmell} of the smells.
We broke \percent[s]{\countBuildError}{\countBuildableDockerfile} of the builds, but it is acceptable for developers that are able to tolerate from \np[\%]{15} to \np[\%]{20} of false positives that developers would tolerate \cite{christakis2016developers}.
We conclude that our approach is suitable for measuring the size impact of the smells.
By side effect, we also show that \tool is effective and could be used by practitioners to detect and fix Docker smells.}

\subsection{RQ2: Impact of Docker Smells}\label{sec:rq3}

\begin{table*}[t]
    \caption{The image size reduction per rule, note that the saving is computed at the image level where several smells could have been repaired. }
    \label{tab:size_reduction}
    \centering
    \begin{tabular}{@{}lrrrrr@{}}
    \toprule
\multirow{2}{*}{Docker Smell} & \multirow{2}{*}{\# Smell} & \multicolumn{4}{c}{Image Size Reduction}   \\\cline{3-6}
  & & Total & Average & Median & Maximum \\
\midrule
aptGetInstallUseNoRec & \np{2242} & \np[GB]{188.1} (6.2\%) & \np[MB]{85.9} (6.2\%) & \np[MB]{17.1} (1.8\%) & \np[GB]{4.4} (87.7\%) \\
pipUseNoCacheDir & \np{2008} & \np[GB]{180.8} (7.1\%) & \np[MB]{92.2} (7.1\%) & \np[MB]{14.6} (1.6\%) & \np[GB]{6.7} (88.3\%) \\
aptGetInstallThenRemoveAptLists & \np{2170} & \np[GB]{163.2} (5.7\%) & \np[MB]{77} (5.7\%) & \np[MB]{13.4} (1.4\%) & \np[GB]{4.4} (87.7\%) \\
npmCacheCleanAfterInstall & \np{1640} & \np[GB]{46.5} (3.6\%) & \np[MB]{29} (3.6\%) & \np[KB]{1.3} (0\%) & \np[GB]{6.7} (88.3\%) \\
tarSomethingRmTheSomething & \np{184} & \np[GB]{4.6} (2.5\%) & \np[MB]{25.7} (2.5\%) & \np[Bytes]{318} (0\%) & \np[MB]{383} (38.7\%) \\
yumInstallRmVarCacheYum & \np{89} & \np[GB]{4.5} (4.8\%) & \np[MB]{51.2} (4.8\%) & \np[Bytes]{177} (0\%) & \np[MB]{801.4} (49.6\%) \\
apkAddUseNoCache & \np{887} & \np[GB]{3.5} (1.5\%) & \np[MB]{4} (1.5\%) & \np[Bytes]{628} (0\%) & \np[MB]{369.5} (32.7\%) \\
mkdirUsrSrcThenRemove & \np{219} & \np[GB]{2.6} (1.1\%) & \np[MB]{12.2} (1.1\%) & \np[Bytes]{275} (0\%) & \np[MB]{205.4} (32.7\%) \\
gemUpdateSystemRmRootGem & \np{34} & \np[MB]{877.4} (2.8\%) & \np[MB]{25.8} (2.8\%) & \np[Bytes]{306} (0\%) & \np[MB]{279.1} (17.6\%) \\
gemUpdateNoDocument & \np{31} & \np[MB]{863.5} (3\%) & \np[MB]{27.9} (3\%) & \np[Bytes]{948} (0\%) & \np[MB]{279.1} (17.6\%) \\
npmCacheCleanUseForce & \np{3} & \np[MB]{56.4} (10.1\%) & \np[MB]{18.8} (10.1\%) & \np[Bytes]{25} (0\%) & \np[MB]{56.4} (18.6\%) \\
rmRecursiveAfterMktempD & \np{2} & \np[Bytes]{43} (0\%) & \np[Bytes]{21.5} (0\%) & \np[Byte]{0} (0\%) & \np[Bytes]{43} (0\%) \\
\midrule
Total & & \totalSpaceSaved (4.66\%) & \avgSpaceSaved & \medSpaceSaved & \maxSpaceSaved \\
\bottomrule
    \end{tabular}
\end{table*}

\begin{table}[t]
    \caption{Impact of the smells on the bandwidth of DockerHub.}
    \label{tab:size_bandwidth}
    \centering
    \begin{tabular}{@{}lrr@{}}
    \toprule
\multirow{2}{*}{Docker Smell}  &  \# Docker Pull& \multirow{1}{*}{Data saved}  \\
  & Per Week & per week \\
\midrule
aptGetInstallUseNoRec & \np{6205156} & \np[TB]{32.76} \\
pipUseNoCacheDir & \np{3591566} & \np[TB]{8.62} \\
aptGetInstallThenRemoveAptLists & \np{3667777} & \np[TB]{12.41} \\
npmCacheCleanAfterInstall & \np{2325059} & \np[TB]{2.94} \\
tarSomethingRmTheSomething & \np{440604} & \np[GB]{42.07} \\
yumInstallRmVarCacheYum & \np{675} & \np[GB]{10.45} \\
apkAddUseNoCache & \np{2048579} & \np[GB]{664.63} \\
mkdirUsrSrcThenRemove & \np{229698} & \np[TB]{1.03} \\
gemUpdateSystemRmRootGem & \np{8675} & \np[GB]{2.64} \\
gemUpdateNoDocument & \np{266} & \np[GB]{2.64} \\
rmRecursiveAfterMktempD & \np{319649} & \np[MB]{4.1} \\
\midrule
Total & \np{\totalPullWeek} & \totalSavedWeek\\
\bottomrule
    \end{tabular}
\end{table}

In this research question, we investigate the impact of Docker smells on the size of Docker images. 
We utilize the \np{\countBuildableDockerfile} buildable Dockerfiles from the previous research question and analyze the differences between the images before and after the repairs. 
Our investigation focuses on image size impact, and on bandwidth usage introduced by the smells.

The results of the size impact investigation are presented in \autoref{tab:size_reduction}. 
The table lists the names of the smells, along with the space used by each smell (difference before and after repair), average used space, median used space, and maximum used space.

It is crucial to acknowledge that while we can observe the space savings per Dockerfile, it is not possible to determine the exact space savings for each smell since we built the Docker images once with all repairs applied.

Overall, the identified smells contribute to an increase in the image size of \totalSpaceSaved (approximately \np[\%]{4.66}).
On average, each Dockerfile exhibits an increase of approximately \avgSpaceSaved in terms of size, with a median of \medSpaceSaved per Dockerfile.

We performed the Wilcoxon signed-rank test to verify if the reduction of size is a significant difference. Wilcoxon signed-rank test is used to compare the locations of two populations using two matched samples and this test is also compatible with non-normal data as is the case here as observed by the Shapiro normality test.
We consider that the reduction in size is significant if the $p-value$ is lower than $0.05$.
We obtained a $p-value$ value of \np{0} which indicates a significant difference in the size before and after the repair.

We also observe a variation in terms of size impact depending on the smell. Some smells, like \smell{npmCacheCleanUseForce}, result in an average impact of approximately \np[\%]{10}.
While other smells like \smell{mkdirUsrSrcThenRemove} only have an impact of \np[\%]{1.1}.
In general, the smells that primarily impact image size are related to package managers, particularly instances where developers forget to remove caches, such as \smell{aptGetInstallThenRemoveAptLists}, \smell{pipUseNoCacheDir}, \smell{npmCacheCleanAfterInstall}, and \smell{aptGetInstallUseNoRec}. These smells are not only among the most frequent but are also relatively straightforward to address.

Additionally, considering the number of times Docker images are downloaded from DockerHub, the impact of these smells becomes more significant. 
\autoref{tab:size_bandwidth} presents the impact of the smells on the bandwidth of DockerHub.
This table only considers the \np{\nbDockerHub} Docker images that we found on DockerHub. \footnote{Collected on January 6th, 2023}
We estimate that the detected smells result in an increase of \totalSavedWeek{} of data transfer per week on DockerHub.
This estimation considers the total downloads for each Docker image and the size difference between the original and repaired Docker images, divided by the median image compression ratio (3.2x) reported by Zhao et al.~\cite{zhao2020large}.

Those numbers can seem non-meaningful for a company the size of DockerHub. 
However, we measured the impact on a small number of images, Dockerhub contains at least used \np{636625} unique images~\cite{opdebeeck2023docker} that are pulled 446 billion times. While considering the full scale of DockerHub those smells have a measurable impact on DockerHub.

\answer{2}{
Docker smells significantly impact the size of Docker images, with an average of \np[\%]{4.66} and going up to \np[\%]{10} for some of the smells. 
This leads to an additional \totalSpaceSavedPerWeek of data transfer per week on DockerHub for \nbDockerHub Docker images.
Among the most frequent and impactful smells we identified, many are related to the use of package managers and their caches. Addressing these smells can have a substantial effect on image size and overall image efficiency.}

\subsection{RQ3: Developers' Attitude Towards Docker Smells}\label{sec:rq4}

\begin{table*}
    \caption{List of the pull requests that receive an answer from the maintainers. The complete list of opened pull requests is available in our repository \cite{repo-expe}.}
    \label{tab:prs}
    \centering
    \setlength\tabcolsep{2.5pt}
    \begin{tabular}{@{}rlrrrr|cc rrr@{}}
    \toprule
\multirow{2}{*}{\#} & \multirow{2}{*}{Project} &\multirow{2}{*}{\# Stars} & \multicolumn{2}{c}{\# Image Pull} & \multirow{2}{*}{Image Size} & \multirow{2}{*}{PR ID} & \multirow{2}{*}{Status} & \multirow{2}{*}{\# Smell} & \multicolumn{2}{c}{Data Saved}  \\\cline{4-5}\cline{10-11}
& &  & Total & Per Week &  &    &  &  & Image & per Week \\
\midrule
1 & \href{https://github.com/AdWerx/pronto-ruby/pull/171}{AdWerx/pronto-ruby} & 20 & \np{198057} & \np{1125} & \np[MB]{857.35} & 171 & Merged & 6 & 38.68 MB (4.51\%) & \np[GB]{13.28} \\
2 & \href{https://github.com/pelias/openaddresses/pull/514}{pelias/openaddresses} & 46 & \np{90188} & \np{304} & \np[MB]{577.31} & 514 & Merged & 2 & 131.65 MB (22.8\%) & \np[GB]{12.2} \\
3 & \href{https://github.com/TomWright/mermaid-server/pull/122}{TomWright/mermaid-server} & 248 & \np{2172} & \np{15} & \np[MB]{889.97} & 122 & Merged & 2 & 28.56 MB (3.21\%) & \np[MB]{136.32} \\
4 & \href{https://github.com/sqlfluff/sqlfluff/pull/4262}{sqlfluff/sqlfluff} & 6876 & \np{43313} & \np{743} & \np[MB]{208.15} & 4262 & Merged & 3 & 11.74 MB (5.64\%) & \np[GB]{2.66} \\
5 & \href{https://github.com/rchakode/realopinsight/pull/30}{rchakode/realopinsight} & 60 & \np{26023} & \np{118} & \np[MB]{809} & 30 & Merged & 2 & 39 MB (4.82\%) & \np[GB]{1.4} \\
6 & \href{https://github.com/vyperlang/vyper/pull/3224}{vyperlang/vyper} & 4695 & \np{72697} & \np{443} & \np[MB]{453.77} & 3224 & Merged & 1 & 23.9 MB (5.27\%) & \np[GB]{3.23} \\
7 & \href{https://github.com/Kruptein/PlanarAlly/pull/1142}{Kruptein/PlanarAlly} & 361 & \np{165010} & \np{848} & \np[MB]{342.55} & 1142 & Merged & 3 & 31.07 MB (9.07\%) & \np[GB]{8.04} \\
8 & \href{https://github.com/ShaneIsrael/fireshare/pull/166}{ShaneIsrael/fireshare} & 522 & \np{10968} & \np{340} & \np[MB]{879.21} & 166 & Merged & 4 & 158.71 MB (18.05\%) & \np[GB]{16.45} \\
9 & \href{https://github.com/jcraigk/kudochest/pull/187}{jcraigk/kudochest} & 18 & \np{2096} & \np{28} & \np[GB]{1.52} & 187 & Merged & 3 & 302.91 MB (19.42\%) & \np[GB]{2.57} \\
10 & \href{https://github.com/fzls/djc_helper/pull/149}{fzls/djc\_helper} & 331 & \np{6682} & \np{95} & \np[MB]{489.2} & 149 & Merged & 4 & 266.1 MB (54.39\%) & \np[GB]{7.68} \\
11 & \href{https://github.com/gotzl/accservermanager/pull/53}{gotzl/accservermanager} & 48 & \np{7040} & \np{35} & \np[GB]{1.14} & 53 & Merged & 1 & 680.38 MB (58.07\%) & \np[GB]{7.32} \\
12 & \href{https://github.com/nitrictech/cli/pull/438}{nitrictech/cli} & 22 & \np{1042} & \np{34} & \np[GB]{1.47} & 438 & Merged & 4 & 113.78 MB (7.55\%) & \np[GB]{1.19} \\
13 & \href{https://github.com/artsy/hokusai/pull/323}{artsy/hokusai} & 89 & \np{396984} & \np{1442} & \np[MB]{539.42} & 323 & Merged & 2 & 10.07 MB (1.87\%) & \np[GB]{4.43} \\
14 & \href{https://github.com/brndnmtthws/tweet-delete/pull/107}{brndnmtthws/tweet-delete} & 92 & \np{14727} & \np{74} & \np[MB]{478.49} & 107 & Merged & 2 & 19.94 MB (4.17\%) & \np[MB]{460.73} \\
15 & \href{https://github.com/bitovi/bitops/pull/390}{bitovi/bitops} & 34 & \np{8496} & \np{70} & \np[MB]{168.45} & 390 & Merged & 2 & 12.31 MB (7.31\%) & \np[MB]{270.08} \\
16 & \href{https://github.com/evennia/evennia/pull/3091}{evennia/evennia} & 1671 & \np{37540} & \np{121} & \np[GB]{1.25} & 3091 & Merged & 5 & 195.49 MB (15.24\%) & \np[GB]{7.22} \\
17 & \href{https://github.com/sbs20/scanservjs/pull/527}{sbs20/scanservjs} & 583 & \np{253233} & \np{1846} & \np[GB]{1.04} & 527 & Merged & 6 & 419.13 MB (39.45\%) & \np[GB]{236.15} \\
18 & \href{https://github.com/mitre/saf/pull/989}{mitre/saf} & 118 & \np{4113} & \np{75} & \np[MB]{603.93} & 989 & Merged & 2 & 124.01 MB (20.53\%) & \np[GB]{2.83} \\
19 & \href{https://github.com/w9jds/firebase-action/pull/176}{w9jds/firebase-action} & 883 & \np{3167517} & \np{28147} & \np[GB]{1.36} & 176 & Merged & 4 & 124.99 MB (8.96\%) & \np[TB]{1.05} \\
20 & \href{https://github.com/naorlivne/terraformize/pull/367}{naorlivne/terraformize} & 151 & \np{9663} & \np{57} & \np[MB]{131.89} & 367 & Merged & 1 & 3.98 MB (3.02\%) & \np[MB]{70.82} \\
21 & \href{https://github.com/nwithan8/tauticord/pull/60}{nwithan8/tauticord} & 78 & \np{1568} & \np{8} & \np[MB]{971.41} & 60 & Merged & 1 & 18.89 MB (1.94\%) & \np[MB]{50} \\
22 & \href{https://github.com/azlux/botamusique/pull/353}{azlux/botamusique} & 290 & \np{174582} & \np{1316} & \np[MB]{667} & 353 & Merged & 2 & 85.4 MB (12.8\%) & \np[GB]{34.31} \\
23 & \href{https://github.com/labsyspharm/scimap/pull/43}{labsyspharm/scimap} & 46 & \np{3456} & \np{44} & \np[GB]{2.15} & 43 & Merged & 1 & 307.74 MB (13.96\%) & \np[GB]{4.11} \\
24 & \href{https://github.com/leighmacdonald/gbans/pull/374}{leighmacdonald/gbans} & 32 & \np{1437} & \np{14} & \np[MB]{40.45} & 374 & Merged & 3 & 2.46 MB (6.09\%) & \np[MB]{10.94} \\
25 & \href{https://github.com/alephdata/aleph/pull/2801}{alephdata/aleph} & 1881 & \np{2254039} & \np{8249} & \np[MB]{990.23} & 2801 & Merged & 4 & 263.57 MB (26.62\%) & \np[GB]{663.54} \\
26 & \href{https://github.com/openedx/credentials/pull/1912}{openedx/credentials} & 20 & \np{1893} & \np{26} & \np[GB]{1.28} & 1912 & Merged & 8 & 132.96 MB (10.15\%) & \np[GB]{1.05} \\
27 & \href{https://github.com/atmoz/sftp/pull/357}{atmoz/sftp} & 1469 & \np{982418661} & \np{2289101} & \np[MB]{155.55} & 357 & Closed & 1 & 26.3 MB (16.91\%) & \np[TB]{17.94} \\
28 & \href{https://github.com/codacy/codacy-eslint/pull/3741}{codacy/codacy-eslint} & 13 & \np{636580} & \np{1685} & \np[GB]{1.38} & 3741 & Closed & 2 & 195.24 MB (13.82\%) & \np[GB]{100.37} \\
\midrule
\multicolumn{8}{@{}l}{\np{\nbPRs} Opened, \percent[p]{\nbAcceptedPR}{\nbPRs} Merged, \percent[p]{\nbRefusedPR}{\nbPRs} Closed, \percent[p]{\nbPendingPR}{\nbPRs} Pending Pull Requests} & \np{\nbMergedRepairedSmell} & \totalSpaceMerged & \totalSpaceSavedPerWeek \\
\bottomrule
    \end{tabular}
\end{table*}


In this final research question, we investigate developers' attitudes toward Docker smells and their impact. 
The main goal is to validate the relevance of these smells to developers and the importance of addressing them. 
To gather feedback from developers, we opened pull requests that addressed Docker smells, and we assessed the impact of these smells within the pull request descriptions.
A template of this description is available in \autoref{lst:pr_template}.

\begin{listing}
    \begin{minted}{text}
Hi there,

I've made a small improvement to the Dockerfile that I think could help optimize the image size.

Summary of the changes:
- <change description>

Impact on the image size:

Image size before repair: <size> MB
Image size after repair: <size> MB
Difference: <size> MB 
I hope that you will find these changes useful to you. Let me know if you have any questions or concerns.

Thanks,
    \end{minted}
    \caption{Template of the pull request description that we used to propose Docker smell repair.}\label{lst:pr_template}
\end{listing}

We established the following criteria to select the repositories where we would submit the pull requests:
\begin{enumerate*}
\item Dockerfile has at least one smell and less than ten.
\item The repository is active (not archived, not a fork, has open issues, at least one fork, and has a commit in the last two months preceding the date of the study on the main branch).
\item The Docker image builds successfully after the repair.
\item No more than one pull request per GitHub organization.
\item The Docker image has been downloaded at least \np{1000} times from Dockerhub.
\item The size difference needs to be larger than \np[Mb]{1}.
\end{enumerate*}
Following these criteria, we identified \np{124} potential candidates and selected \np{\nbPRs} repositories that explicitly welcome external contributions. The list of opened pull requests can be found in our repository \cite{repo-expe}.

The merged and closed pull requests are presented in \autoref{tab:prs}. The table includes the repository name, the number of stars, and the total and average weekly downloads on Dockerhub. Additionally, it shows the original image size, pull request ID, pull request status, number of repaired smells, image size reduction, and the theoretical average bandwidth saving per week (considering a median compression rate of \np{3.2} \cite{zhao2020large}).

Out of the \np{\nbPRs} pull requests, \percent[p]{\nbAcceptedPR}{\nbPRs}, were accepted and merged successfully including one required manual change.
The remaining \percent[p]{\nbPendingPR}{\nbPRs} pull requests are awaiting responses from the developers. The accepted pull requests resulted in a total saving of \totalSpaceMerged, which translates to a weekly saving of \totalSpaceSavedPerWeek, considering the \np{\mergedDownloadTimeWeek} average weekly downloads.

Some pull requests triggered some discussions; while other pull requests were simply merged by developers without interaction.
But most developers simply appreciated the contribution and merged the pull requests, as seen in \href{https://github.com/Kruptein/PlanarAlly/pull/1142}{PR-7} and \href{https://github.com/jcraigk/kudochest/pull/187}{PR-9}. While those feedbacks do not explicitly address the Docker smells, it does indicate that developers value such contributions, suggesting they consider Docker smells as relevant. 
In a different case, developers explicitly expressed their interest in the changes, as seen in \href{https://github.com/gotzl/accservermanager/pull/53}{PR-11}: \textit{Hi, thank you very much, this change seems quite sensible! Cheers}. 

\begin{figure}
    \centering
    \includegraphics[width=0.95\linewidth]{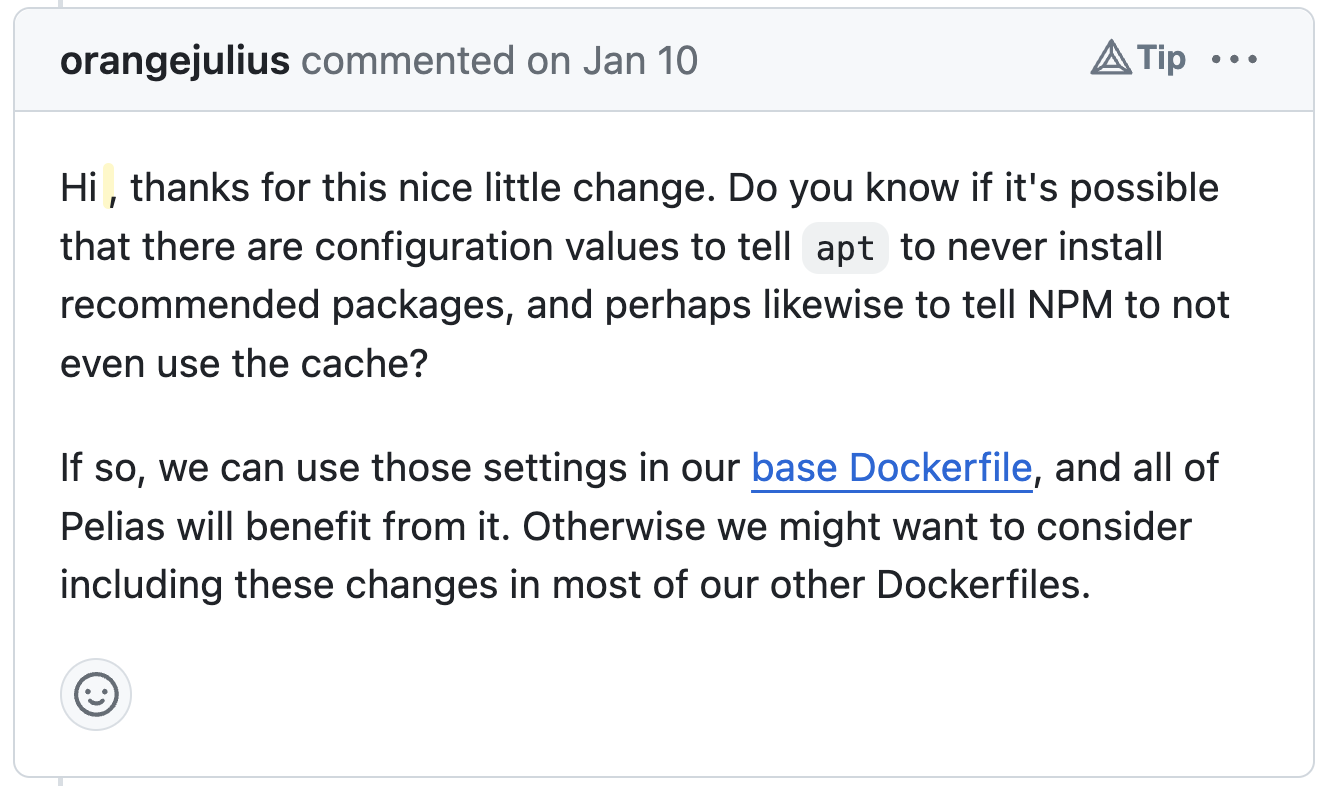}
    \includegraphics[width=0.95\linewidth]{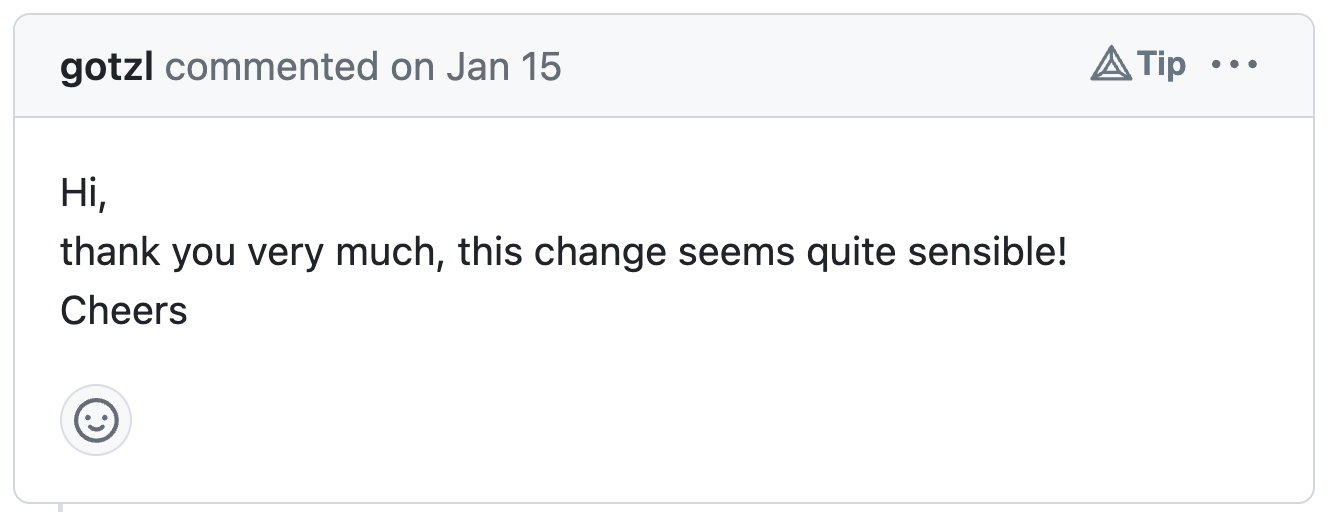}
    \includegraphics[width=0.95\linewidth]{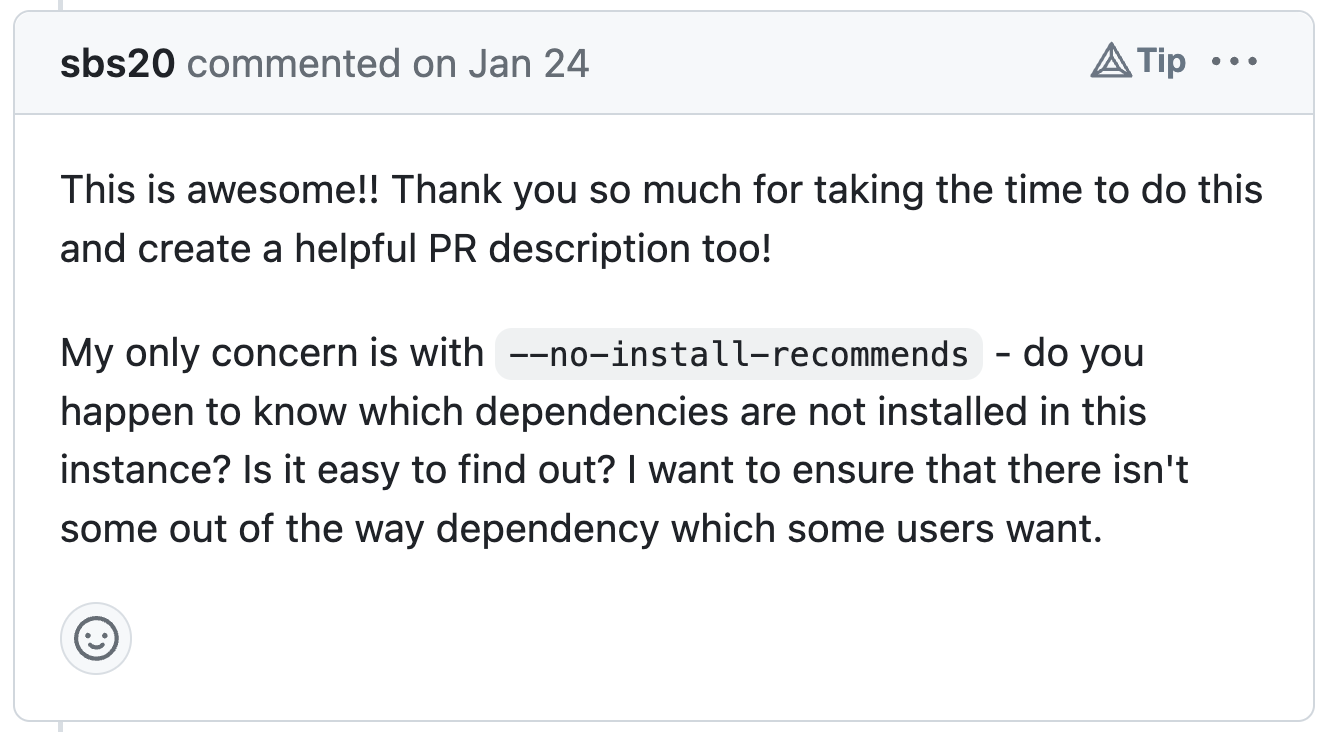}
    \caption{Comment examples of merged PRs: \href{https://github.com/pelias/openaddresses/pull/514}{PR-2},\href{https://github.com/gotzl/accservermanager/pull/53}{PR-11},\href{https://github.com/sbs20/scanservjs/pull/527}{PR-17}}
    \label{fig:accepted-comments}
\end{figure}

Some other pull requests triggered additional discussions as illustrated in \autoref{fig:accepted-comments}.
The developers asked if you could apply the changes to the base image of their application as seen in \href{https://github.com/mitre/saf/pull/989}{PR-18}, some other repositories wanted to know about the tool that we use to create the fix such as \href{https://github.com/pelias/openaddresses/pull/514}{PR-2}.

During the discussions, there were instances of developers expressing concerns about specific repairs, notably regarding the \smell{aptGetInstallUseNoRec} smell, which pertains to not installing recommended packages. For instance, developers in \href{https://github.com/sbs20/scanservjs/pull/527}{PR-17} were worried about the impact and asked us to remove that part of the change. Nonetheless, they appreciated the contribution and expressed gratitude for learning something new about Docker and \code{apt}.

\begin{figure}
    \centering
    \includegraphics[width=0.95\linewidth]{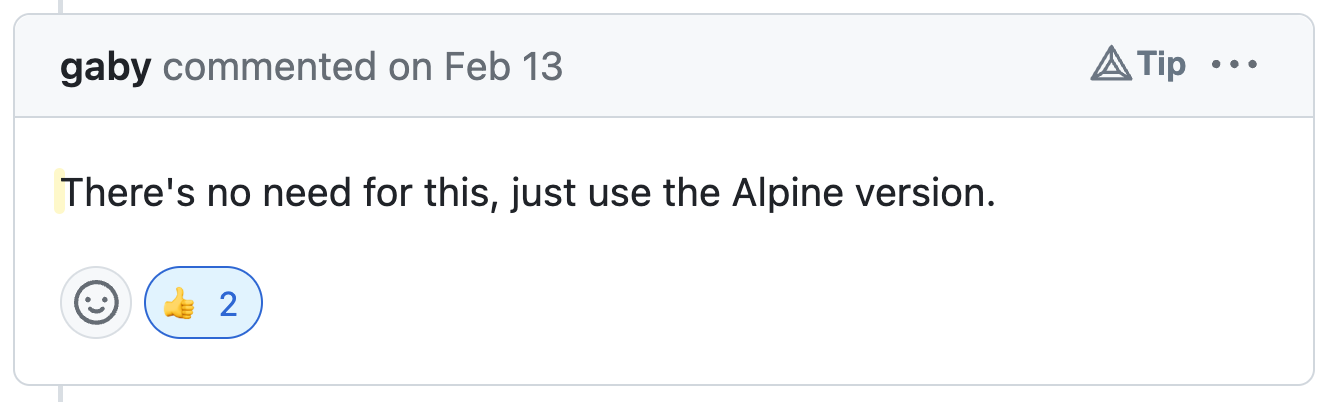}
    \includegraphics[width=0.95\linewidth]{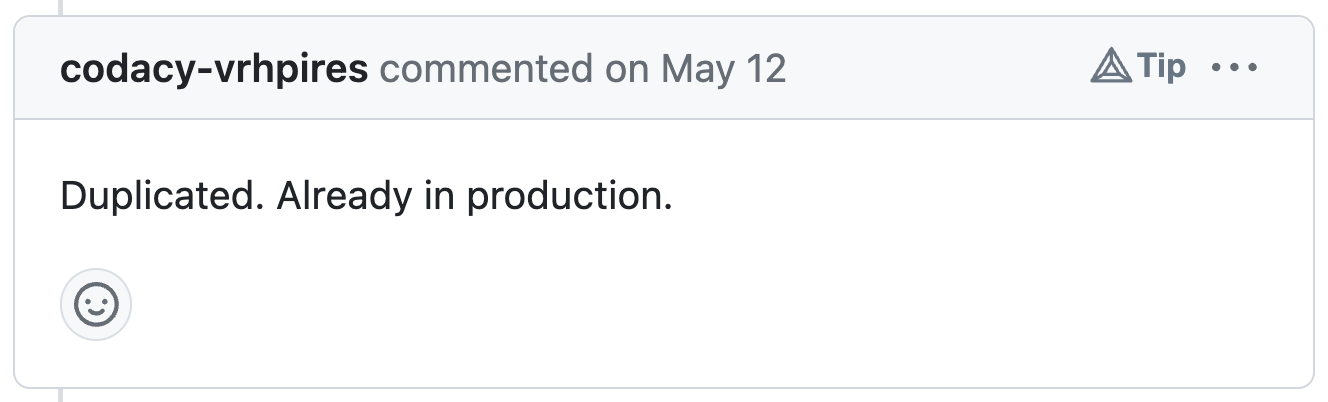}
    \caption{Comments of rejected PRs: \href{https://github.com/atmoz/sftp/pull/357}{PR-27},\href{https://github.com/codacy/codacy-eslint/pull/3741}{PR-28}}
    \label{fig:rejected-comments}
\end{figure}

Regarding the pull requests that were rejected, developers informed us that the changes were already present in the production or Alpine version as illustrated in \autoref{fig:rejected-comments}. In \href{https://github.com/atmoz/sftp/pull/357}{PR-27}, the maintainer said \textit{There's no need for this, just use the Alpine version.}, and in \href{https://github.com/codacy/codacy-eslint/pull/3741}{PR-28}, they said \textit{Duplicated. Already in production.}. While our pull requests were not merged, the fact that the changes were already implemented indicates that the smells are still considered relevant.

An important outcome of this research question is the acceptance by the developers. We have not faced yet a case where the developers do not find the change relevant.
This is an interesting result compared to how smells are generally considered by developers.
We expect that the perspective of the developers is different in this case because the impact of the smell can be directly measured, the number of false positives is low and the fix is comprehensive.
This observation motivates us to extend further the ability of \tool and to propose automatic patches for the Docker Smells.


\answer{3}{We submitted \np{\nbPRs} pull requests, \np{\nbAcceptedPR} have been accepted, two have been rejected.
Overall, the merged pull requests and the feedback that we received from the developers are overwhelmingly positive where developers acknowledged the fix even in the case of rejected pull requests. This observation contrasts with the general treatment that code smells are receiving from developers which highlights the importance and relevance of this study.}

\section{Related work}

Docker has become a popular tool for developers and organizations to package, deploy, and run applications in a lightweight, portable container. As such, there has been a significant amount of research focused on improving the efficiency, security, and maintainability of Docker-based projects. In this section, we review several relevant studies that are related to this contribution.

A large number of papers studied the Docker ecosystem. We present a selection of them.
Ibrahim et al. \cite{IbrahimSH20} investigate the number and diversity of images available on DockerHub for the same system, finding that there is a large number of images to choose from and significant differences between them. 
Ksontini et al. \cite{ksontini2021refactorings} study the occurrence of refactorings and technical debt in Docker projects, finding that refactorings are common but technical debt is rare.
Xu et al. \cite{xu2018mining} present a study of mining container image repositories for software configuration information, finding that such information is often incomplete or outdated.
Lin et al. \cite{lin2020large} study the Docker images hosted on DockerHub. They observe a downward trend of Docker image sizes and smells in Dockerfiles.
However, they also observed an upward trend in using obsolete base images.
Lui et al. \cite{liu2020understanding} also study DockerHub but focused on the security risks associated with it.
Eng et al. \cite{EngH21} did a longitudinal study of the evolution of Dockerfiles, and they confirm that there are slightly fewer smells over time.
However, none of those papers study the impact of the smells on the Docker image size.

Other works also focus on improving the security of containers, such as
SPEAKER \cite{lei2017speaker}, which reduces the number of available system calls to a given application container by customizing and differentiating its necessary system calls at the booting and the running phases.
Confine \cite{ghavamnia2020confine} is a similar technique that uses static analysis to identify the required system calls. 

Other contributions aim to improve or fix Dockerfiles.
Henkel et al. \cite{HenkelSTdR21} propose an approach for repairing Dockerfiles that do not build correctly. It uses machine learning to infer repair rules based on build log analysis. 
Hassan et al. \cite{hassan2018rudsea} present Rudsea, a technique that adapts Dockerfiles based on the changes in the rest of the project.
Zhang et al. \cite{zhang2020recommending} propose a technique that recommends Docker base images to improve efficiency and maintainability.
Other tools aim to reduce the size of the Docker images by identifying bloat in the images and removing it.
Cimplifier \cite{rastogi2017cimplifier} and their framework \cite{rastogi2017new} aim to automatically partition containers into simpler containers based on user-defined constraints. The goals are isolation of each sub-container, communicating as necessary, and only including enough resources to perform their functionality.
strip-docker-image \cite{strip-docker-image}, minicon \cite{minicon} and docker-slim \cite{SlimToolkit} are open-source projects that reduce Docker image size by specializing the container to the application.

An important part of the bloat comes from bad practices. Several tools and works focus on identifying those Docker smells.
Binnacle \cite{10.1145/3377811.3380406} is a tool for detecting Docker smells, they compared the presence of those smells between a set of Dockerfiles from GitHub and a set of Dockerfiles written by experts.
They observed that there are five times fewer smells in the export Dockerfiles.
Wu et al. \cite{wu2020characterizing} study the docker smell occurrence in \np{6334} projects. They show that smells are very common and there exists co-occurrence between different smells.
Xu et al. \cite{xu2019dockerfile} propose a technique based on static and dynamic analysis to detect temporary files inside Dockerfiles.
Nonacademic works focus on detecting Dockerfile smells: Hadolint \cite{hadolint}, dockerfilelint \cite{dockerfilelint}, docker-bench-security \cite{docker-bench-security}, or dockle \cite{dockle}.
However, none of these tools aim to repair the detected smells or analyze the impact of those smells.

Overall, there has been a significant amount of research focused on Docker, including tools for debloating, optimizing, and securing containers, as well as studies of the evolution and management of Dockerfiles and images. 
However, this empirical study is as far as we know the first that studies the impact of the smells on the Docker images and that collects feedback from the developers.

\section{Threats to Validity}

In this section, we explore potential threats to the validity of our study and detail the measures taken to address them, thereby bolstering confidence in our results. Our classification framework aligns with the model proposed by Wohlin et al. \cite{wohlin2012experimentation}.

\subsection{Construct Validity}

Construct validity threats stem from the alignment between theory and observation, largely influenced by the measurement procedures in our study. To address this, we took a meticulous approach. Firstly, we selected smell types reported and measured by a different research group. These smells were presented to practitioners, and their impact was measured. Another potential threat arises from the study's limited scope, focusing on specific smells in bash Dockerfiles. We mitigated this by verifying the presence of smells in recent Dockerfiles and presenting them to developers through pull requests.

Additionally, our focus on bash Dockerfiles excludes those in PowerShell, but given the prevalence of bash in Dockerfiles, our results remain relevant for the majority of users. We aimed to eliminate potential bias or subjectivity in the technique selection process.

\subsection{Internal Validity}

Internal validity focuses on establishing a reliable causal relationship between a treatment or intervention and its observed outcomes. One potential threat is the presence of internal bugs in \tool. To address this, extensive testing was conducted, and \tool was made open-source, enabling scrutiny by developers and researchers. Another potential threat involves the diversity of our dataset. To mitigate this, we collected a large and diverse dataset of Dockerfiles and supplemented our pull request selection with an existing dataset (Binnacle).

\subsection{External Validity}

External validity concerns the generalizability of study results. To enhance external validity, experiments were conducted on diverse case studies from different open-source projects, spanning various languages and sizes. While we focused on measuring the impacts of smells in terms of size and bandwidth on Docker images, this might limit the generalization of our results to all smell types. However, this specific impact aligns with common effects of smells, as supported by \cite{10.1145/3377811.3380406}. The relevance of image size in distributed systems further strengthens the importance of considering size increases in the evaluation of distributed software.

\subsection{Conclusion Validity}

Threats to conclusion validity involve the connection between the treatment and outcome, specifically regarding the reproducibility of the study's findings. To address this concern, we conducted experiments with a rigorous and mostly automated methodology. We also evaluated the precision and recall of the tool used in the study to ensure that our observations are reproducible. This comprehensive approach provides ample evidence to draw valid conclusions. Moreover, to ensure replicability, a rigorous methodology was followed in performing the experiments. The source code, scripts, and procedures are thoroughly documented, enabling other researchers to replicate the study with precision.

\section{Conclusion}

In this paper, we present an empirical study of the impact of Docker smells on image size.
For this study, we identify and repair \np{\nbRepairedSmellInBuildable} Docker smells from \np{\nbSmellInBuildable} Dockerfiles.

We observe that smells lead to an average increase in image size by \np[\%]{4.66} and a total of \totalSavedWeek of transfer per week (on DockerHub).
Interestingly, the most common smells are related to the package managers and they are the smells that impact the most the image size.

Additionally, we verify the relevance of the smells by opening \np{\nbPRs} pull requests on open-source projects that fix the Docker smells and reduce the Docker image size. 
We found that the developers react overwhelmingly positively to the pull requests by merging \percent[p]{\nbAcceptedPR}{\nbPRs} and by providing feedback that confirms that the smells are relevant to them even in the two cases where our pull requests were rejected.

The detection and repair of the smells has been performed by our tool \tool. This study consequently also highlights the relevance of such a tool to help practitioners improve their Dockerfiles and therefore their Docker image.


Those results motivate us to continue this line of research and improve the Docker ecosystem. In particular, we aim to extend \tool to support additional smell, integrate it inside IDE for direct developer feedback and also work on estimating the impact of the smells without having to build the Docker images.

\section*{Data Availability}
We provide the scripts, dataset, and tool used in this contribution. 
You can find \tool at \cite{repo}, and the empirical study data at \cite{repo-expe} as well as a functional demo of \tool at \url{\urlDemo}.

\bibliography{references}
\end{document}